\documentclass[sigconf]{acmart}
\usepackage{amsmath,amssymb,amsthm,amsfonts}
\usepackage{algorithmic}
\usepackage{graphicx}
\usepackage{textcomp}
\usepackage{xcolor}
\usepackage{tikz,color,subcaption,url, comment}
\usepackage{multirow}
\usepackage{diagbox}
\usepackage{verbatim}
\usepackage{eucal}

\AtBeginDocument{%
  \providecommand\BibTeX{{%
    \normalfont B\kern-0.5em{\scshape i\kern-0.25em b}\kern-0.8em\TeX}}}

\setcopyright{acmcopyright}
\copyrightyear{2018}
\acmYear{2018}
\acmDOI{10.1145/1122445.1122456}

\acmConference[MSWiM '19]{MSWiM '19: The 22nd ACM International Conference on Modeling, Analysis
and Simulation of Wireless and Mobile Systems}{November 25th - 29th, 2019}{Miami}
\acmBooktitle{ }
\acmPrice{ }
\acmISBN{ }

\begin{document}

\title{Markovian model for Broadcast in Wireless Body Area Networks}


\author{Bruno Baynat}
\affiliation{%
  \institution{Sorbonne University, LIP6, CNRS UMR 7606\\
  Email: Bruno.Baynat@lip6.fr}
  \city{Paris}
  \country{France}
}

\author{Gewu Bu}
\affiliation{%
  \institution{Sorbonne University, LIP6, CNRS UMR 7606\\
  Email: Gewu.Bu@lip6.fr}
  \city{Paris}
  \country{France}
}

\author{Maria Potop-Butucaru}
\affiliation{%
  \institution{Sorbonne University, LIP6, CNRS UMR 7606\\
  Email: maria.potop-butucaru@lip6.fr}
  \city{Paris}
  \country{France}
}

\renewcommand{\shortauthors}{Baynat et al.}

\begin{abstract}
Wireless body area networks became recently a vast field of investigation. 
A large amount of research in this field is dedicated to the evaluation of various communication protocols, e.g., broadcast or convergecast, against human body mobility.
Most of the time this evaluation is done via simulations and in many situations only synthetic data is used for the human body mobility.  
In this paper we propose for the first time in Wireless Body Area Networks a Markovian  analytical model  specifically designed for WBAN networks.  The main objective of the model is to evaluate the efficiency of a multi-hop transmission in the case of a diffusion-based broadcast protocol, with respect to various performance parameters (e.g., cover probability, average cover number, hitting probability or average cover time).
%with and without interferences. In order to validate the accuracy of our model, we use as case study a diffusion-based broadcast protocol. Our model allows to compute various performance parameters (e.g., cover probability, average cover number, hitting probability or average cover time) of the protocol.
We validate our model by comparing its results to simulation and show its accuracy. 
%Furthermore, we compare the accuracy of the performance derived from our model with simulations. Interestingly, in the case of the cover probability the average relative error between the model and the simulation is 14\% for the non-interference case and only 6\% for the interference case.
Finally, but not least, we show how our model can be used to analytically evaluate the trade-off between transmission power and redundancy, when the same message is broadcasted several times in order to increase the broadcast reliability while maintaining a low transmission power.
\end{abstract}


%\ccsdesc[300]{Networks~Peer-to-peer protocols}

 % \includegraphics[width=\textwidth]{sampleteaser}
  %\caption{Seattle Mariners at Spring Training, 2010.}
  %\Description{Enjoying the baseball game from the third-base
  %seats. Ichiro Suzuki preparing to bat.}
  %\label{fig:teaser}

\maketitle

\section{Introduction}
 WBAN (Wireless Body Area Networks) (\cite{DBLP:journals/corr/abs-1303-2062},\cite{Latre:2011:SWB},\cite{WBAN1}) are a viable solution in response to various disadvantages associated with wired sensors commonly used to monitor patients in hospitals and emergency rooms. Recent medical reports predict that the number of people using  health technologies will drastically increase from 14.3 to 78 million consumers from 2014 to 2020~\cite{32}, respectively.  
 
\emph{Wireless Body Area Network}, WBAN is a  dedicated wireless sensor network where  tiny devices with low computing power and limited battery life are placed on human body in order to collect various physiological information (e.g., EEG, blood pressure, body temperature, etc.), and further transmit it to a collector point (called the \emph{Sink}) that will process, take decisions, alert or record. This process is called \emph{convergecast}. The sink node can play a controller role and  send system configurations to other WBAN devices to allow them to work together. For example, sink can send \emph{time slots} information for enabling a \emph{Time Division Multiple Access (TDMA)} communication protocol, or send \emph{Transmission Power Modification Requirement} for a \emph{Power Control} mechanism. In this case the sink triggers a \emph{broadcast} transmission.
%Note that these system configuration information is usually small enough to be put into one packet. 

WBANs differ from typical large-scale wireless sensor networks in many aspects \cite{6755575}, \cite{islam2015internet}, \cite{salayma2017wireless}. The size of the network is limited to a dozen nodes, in-network mobility follows the body movements. That is, links have a very short range and a quality that varies with the wearer's posture, but remains low in general.  Moreover, in WBAN  the wireless channel has its specificities. Indeed, the transmission power is kept low, which improves devices autonomy and reduces wearers electromagnetic exposition. Consequently, the effects of body absorption, reflections and interference cannot be neglected and it is difficult to maintain a direct link (one-hop) between a data collection point and all WBAN nodes.

Although, recent research \cite{naganawa2015simulation}, \cite{liu2016performance} advocates for using \emph{multi-hop} communication in WBAN, very few multi-hop communication protocols have been proposed so far and even fewer are optimized for the human body mobility \cite{badreddine:hal-01469314}, \cite{BU201828}. 
Multi-hop convergecast has been studied in the context of WBANs in \cite{badreddine2017convergecast} and most of the proposed solutions have been evaluated using synthetic models for human body. 
Recently, in \cite{BU2018200} the authors propose new strategies and  evaluate them against realistic human body mobility models. 
 Multi-hop broadcast has been studied for the first time in WBAN in \cite{badreddine2015broadcast} and \cite{badreddine:hal-01404848}. The authors propose  a set of strategies and extensively evaluate  them via simulations using OMNET++ simulator and the Mixim framework \cite{mixim}. The most up to date state of the art on broadcast and convergecast in multi-hop settings can be found in \cite{phdthesis}. None of the previously mentioned works proposes a formal model of wireless communication in WBAN neither a formal analysis of the proposed protocols. 
 
 Interestingly, even though simulations, especially in the WBAN context, offer interesting insights on the behavior of various protocols face to human body mobility, it is difficult to tune manually 
 their parameters using only simulation. The objective of this paper is to offer an analytical  framework that allows to automatically  fine tune various parameters of WBAN communication protocols  in order to obtain the best tradeoff between their reliability and the used transmission power.   

\paragraph{Our contribution.}
Our contribution is twofold.
\emph{First}, we propose for the first time in Wireless Body Area Networks an analytical model specifically designed for WBAN networks.
We use as case study a diffusion-based broadcast protocol where a specific node called \emph{sink} has to send some setup information to all other nodes in the network. The protocol is \emph{multi-hop}, i.e., any node in the WBAN may act as a relay for retransmitting a given packet.
We start by developing a simple Markovian model that neglects the possible simultaneous transmission of the same broadcasted packet. Then we extend the model by introducing the effect of interferences between overlapping transmissions. The objective of both models is to evaluate the efficiency of the multi-hop transmission with respect to various performance parameters such as: 
%The main objective of the model is to evaluate the efficiency of the multi-hop transmission with and without interferences. In order to validate the accuracy of our model, we use as case study a diffusion-based broadcast protocol where a single packet is transmitted by a specific node called \emph{sink} and has to be received by all nodes in the network. Our model allows to compute various performance parameters of the protocol such as
\emph{cover probability} (i.e., the probability that all non-sink nodes receive successfully the transmitted packet), \emph{average cover number} (i.e., the average number of non-sink nodes that receive successfully the transmitted packet), \emph{hitting probability} (i.e., the probability that a selected non-sink node receives successfully the consider packet) or \emph{average cover time} (i.e., the average time necessary for all non-sink nodes to receive successfully the considered packet conditioned by the fact that the broadcast is a success).
%We further compare the accuracy of the performances derived from our model with simulations of the protocol  
We validate our model by comparing its results to simulations using OMNeT++ simulator enriched with Mixim framework, where realistic human body mobility data sets have been used to implement the wireless channel.
%\emph{Second}, we use our model  to analytically evaluate the trade-off between transmission power and redundancy when the same message is broadcasted several times in order to increase the broadcast reliability while maintaining a low transmission power.
\emph{Second}, we use our model to evaluate a multi-broadcast strategy that consists in broadcasting several times the same packet. The objective is to increase the broadcast reliability while maintaining a low transmission power. The simplicity of our model enables to analytically quantify the trade-off between transmission power and redundancy and allows us to derive Erlang-like abaques in order to dimension the parameters of the protocol.

%in order to provide a highly reliable broadcast while keeping low transmission power, we have proposed a multi-broadcast protocol that consists in broadcasting several times the same packet. The simplicity of our model has allowed us to evaluate the approach and to show the trade-off between transmission power and redundancy. In future work, we aim at showing that, even though our model is dedicated to broadcast, it can be easily adapted to other communication protocols in WBANs, such as the convergecast of data collection. 

\paragraph{Paper Roadmap.}
Section \ref{System} introduces the WBAN network and our case study. In Section \ref{Model} we first present our model in a simplified version, then we propose a refinement of the model that captures network interferences. In Section \ref{NumRes} we validate our model by comparison with simulation and use it to evaluate the multi-broadcast strategy.

\section{Broadcast in Wireless Body Area Network}
\label{System}

\subsection{Body Area Network Scenario}
\begin{figure}
\centering
\includegraphics[width=0.48\textwidth]{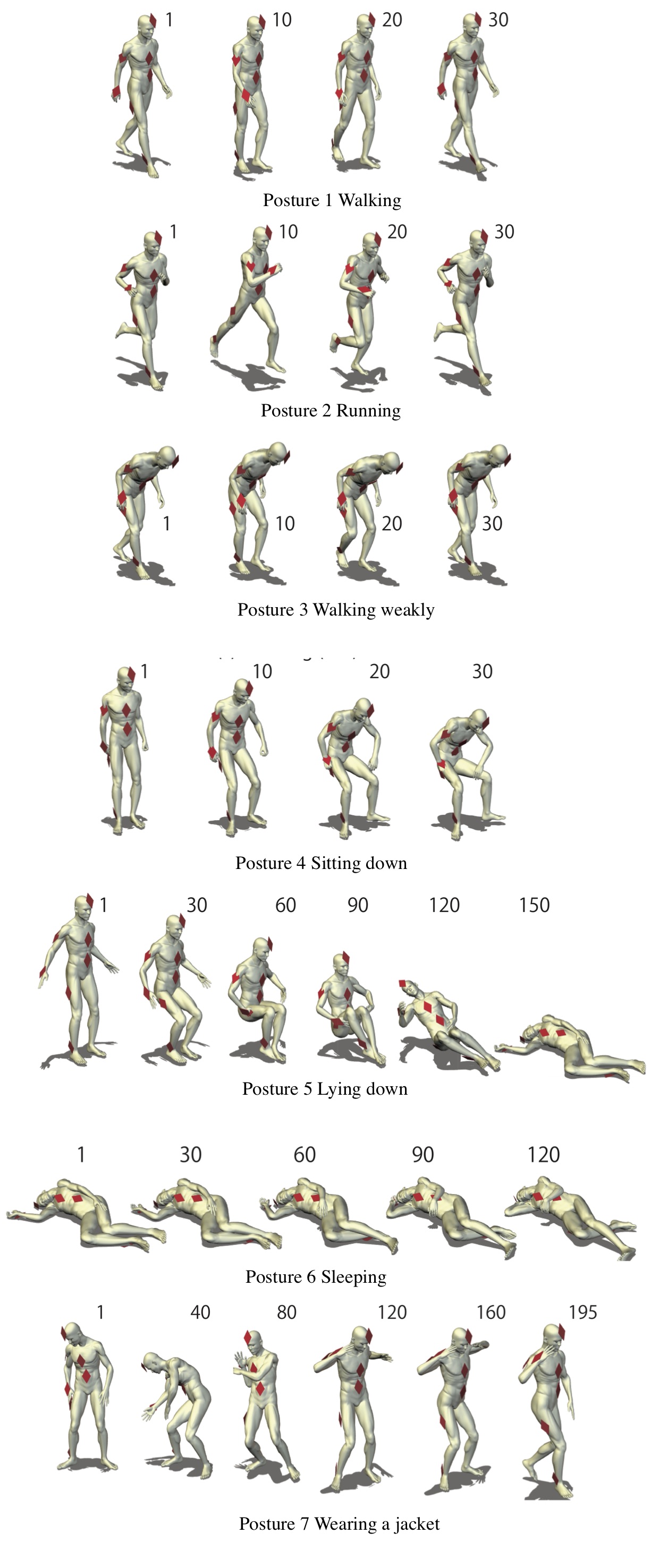}
\caption{7 Different Human Postures \cite{naganawa2015simulation}}
\label{walk1}
\end{figure}

We use the WBAN scenario proposed in  %\cite{badreddine2015broadcast} based on the data sets of 
\cite{naganawa2015simulation}: a WBAN system of seven devices distributed on the body as follows: 0) navel, 1) chest, 2) head, 3) upper arm, 4) ankle, 5) thigh and 6) wrist, shown in Figure \ref{walk1} (the red diamonds in the figure represent WBAN devices on the human body). Node 1 is the Sink (the node that will act as a controller). Using a software-simulated-human-body  authors of \cite{naganawa2015simulation} measure the mean and the standard deviation of the channel attenuation in between each pair of nodes in seven different postures: 1) walking, 2) running, 3) walking weakly, 4) sitting down, 5) lying down, 6) sleeping and 7) putting on a jacket, respectively (see Figure \ref{walk1}). In each posture, a continuous human action has been decomposed into a set of frames. Each single human body picture with a corresponding frame number $x$, is a screenshot of this continuous  action at the $x$th frame. For example, in posture 1 (see Figure \ref{walk1}), the continuous action takes 30 frames. The figure shows four screenshots: 1st frame, 10th frame, 20th frame and the 30th frame, respectively to represent this action. 

%\subsection{Broadcast}
%\subsection{} 
We focus a simple \emph{Broadcast algorithm}  where the sink has to transmit control packets to all non-sink nodes. For a given packet, the algorithm is as follows:
%The sink will send the configuration requirement included into one size-fixed packet with $N_{bit}$ bits to others nodes. Described as follows: 
\begin{itemize}
\item Initially the sink sends its packet into the wireless channel.%. It will send no more packet during the broadcast.
\item A non-sink node that  successfully receives the packet for the first time, will re-send this same packet only once.
\end{itemize}
The broadcast of this considered packet is considered as finished if no more node in the network sends or receives this packet or a copy of it. The broadcast is considered as a success if after its termination all non-sink nodes have successfully received (directly or indirectly) the considered packet.

 %shows an illustration of the broadcast in WBAN.

%\begin{figure}
%\centering
%\includegraphics[width=0.48\textwidth]{over.jpeg}
%\caption{Sink first waits  a random time, in (1). Then, it sends a packet to the wireless channel, in (2). Non-sink nodes  receive the packet successfully, in (3) and begin to re-send it, in (4).}
%\label{over}
%\end{figure}

In order to formally model the behavior of this protocol in a WBAN network we discuss in the following 
some particularities of the wireless communication in WBAN altogether with the way the interferences are handled.

\subsection{Wireless Communication in WBAN}
The connections between nodes are wireless in WBAN with operating frequency bands varying from $400MHz$ to $2400MHz$. The most common band are between $2400MHz$ and $2483.5MHz$, called $2.4G$ band. %Collected data will be transformed into a wireless signal for further wireless transmission. 
%By  its \emph{Analog to Digital Conversion (ADC)} module, a WBAN device transforms collected data from human body to a string of digital binary symbols. These symbols altogether  with some additional control information form a \emph{packet}. Though modulation,  a packet is integrated into a \emph{carrier signal} and is ready to send.
%. So that a signal containing a complete packet is ready to send.
% Note that sending a packet to the wireless channel needs a transmission time, $t_t$. 
%The communication in WBAN is based on these wireless connexions between nodes. 
%\subsubsection{Transmission Power}
Each node can transmit with some \emph{transmission power} denoted as $PT$ in the rest of the paper, that defines the initial strength of the wireless signal. 
%The choice of this $pw_t$ should ensure a sent signal is "strong enough", so that it can be received by at least some other nodes in WBAN system. Note that according to the nature of wireless transmission, a signal is a wireless wave, that could be received by all nodes hearing this signal, if it is "strong enough". Otherwise, under the influence of wireless channel attenuation (we talk about this later), nodes with small $pw_t$ may not communicate with others, they will be isolated from others and therefore can no longer be seen as a part of the WBAN system.
%\subsubsection{Sensitivity}
%What's that mean "strong enough" for a signal that can be heard by others nodes?

A wireless signal sent from a node $u$ can be heard by a node $v$ if the signal strength associated with each bit of the transmission and received by $v$ is higher than the \emph{sensitivity} of $v$.
%Note that the smaller the sensitivity a receiver has, the feebler signal it can detect. 
In other words, signals received by a node with a reception power $PR$ that is smaller than the sensitivity $SN$ be ignored by the receiver.
%\subsubsection{Channel Attenuation}

%$PT$ and $PR$ are not  identical in  WBAN communications. 
During the wireless transmission, the signal strength will be attenuated by the transmission medium which in the case of WBAN is the human body. By the reflection, the refraction and the absorption of human body, the reception power of a signal  is usually different than transmission power because of the randomness of the \emph{Wireless Channel Attenuation} in between the sender and the receiver. During transmission, each bit of the packet will be affected by different attenuation values. 
A transmission is considered as being received successfully only if the packet included in the signal is demodulated correctly at the receiver. 
%\subsubsection{Bit Error Rate}
As the channel is not perfect, packets going through the channel have always a certain loss. Depending on the chosen modulation/demodulation, the lossy probability, called \emph{Bit Error Rate (BER)}, can be computed as a function of $PR$ and the \emph{environmental interference} to this signal. The $BER$ is the probability that a bit of a packet is received wrongly.  The receiver can check if the arriving packet is correct using \emph{checksum} (a string of binary symbols, additional control information added to the packet). Even if only one bit of packet is wrong, the whole packet will be considered as \emph{broken}  and rejected by the receiver. Note that  \emph{checksum} help in detecting the errors but cannot be used to correct them. Note also that, the stronger the $PR$ of a signal, the smaller the $BER$ (provided that the environmental interference and the modulation mode are fixed). A signal is received successfully by the receiver if the $PR$ of this signal is above the sensitivity of the receiver and, according to the $BER$, all bits of the packet  are demodulated correctly.
 
%\subsubsection{Handling Interferences}%collision de ??????????zvi????????????collision???????collision???CSMA-CA (??????????, ????collision???????)

\subsection{Handling Interferences in WBAN}
%In the previous section, we mentioned that BER for a given modulation is  function of receiving signal strength $PR$ and environmental interference to the target signal. $PR$ is impacted by the transmission power $PT$ and the channel attenuation $A$. 
There are two kinds of environmental interference in a WBAN system: one is the \emph{environmental noise} the other one is the \emph{intra-BAN interference}. The environmental noise is inevitable and may always disrupt the reception of a signal. The intra-BAN interference to a target signal at the receiver is made of a set of non-target signals present during the reception of the target signal. %See Figure \ref{interference} for an illustration of internal interference between two signals. 
%The more total strength the internal interference holds, the harder a receiver can decode the target signal successfully. Because the target will be submerged into these interference signals. Note that, the more signals sent into the WBAN in short time interval, the more internal interference may occur at the receiver.  

%\begin{figure}
%\centering
%\includegraphics[width=0.48\textwidth]{interference.jpeg}
%\caption{Two signal $1$ and $2$ are sent into the channel. The two rectangles represent the transmission time of transmitted signals. At time $t0$, the first bit of signal $1$ arrives at the receiver. During the time period $[t0, t1]$, the reception of signal $1$ is not interferenced by signal $2$. However in the time period $[t1, t2]$, signal $2$ also arrives at the receiver. So in this period, the reception of signal $1$ will be affected by signal $2$. Note that even though signal $1$ is received first and during $[t2,t3]$ signal $2$ has no interference, signal $1$  cannot be received because signal $2$ has already been considered as interference during the reception of signal $1$.}
%\label{interference}
%\end{figure}

 In order  to reduce the interference influences during signal reception, an \emph{Unslotted Carrier Sense Multiple Access / Collision Avoid (CSMA/CA)} algorithm is used in WBAN. The idea of CSMA/CA algorithm is to let nodes that have packets to send wait for some random time before sending their packets into the channel. %So that the transmissions of packets could be distributed to different time periods to avoid the presence of multi transmissions in short time interval. 
A brief description of CSMA/CA is as follows:
\begin{enumerate}
\item Before sending a packet to the wireless channel, a node chooses a random number $r$ from an interval $I$ between $0$ and $2^W-1$. Initially, $W$ is set to $3$. 
\item The node launches a timer and waits $r \times T_u$, where $T_u$ is a fixed time unit, called  \emph{backoff unit}.  $T_u = 0.32$ ms according to the standard 802.15.4 specification for CSMA/CA.
\item When the timer expires and after a configuration time, the node listens to the wireless channel. If the channel is busy, which means that there are other signals under the transmission in the channel, then the node goes back to step 1) with $W=W+1$. Otherwise,  the node begins to send its packet through the channel.
\end{enumerate}
 The maximal number of times a node can re-arm the backoff timer is $Max$.  The standard 802.15.4 for CSMA/CA recommends setting  $Max=5$.  
 Passed this limit, the waiting packet will be discarded.
 
\begin{figure}
\centering
\includegraphics[width=0.48\textwidth]{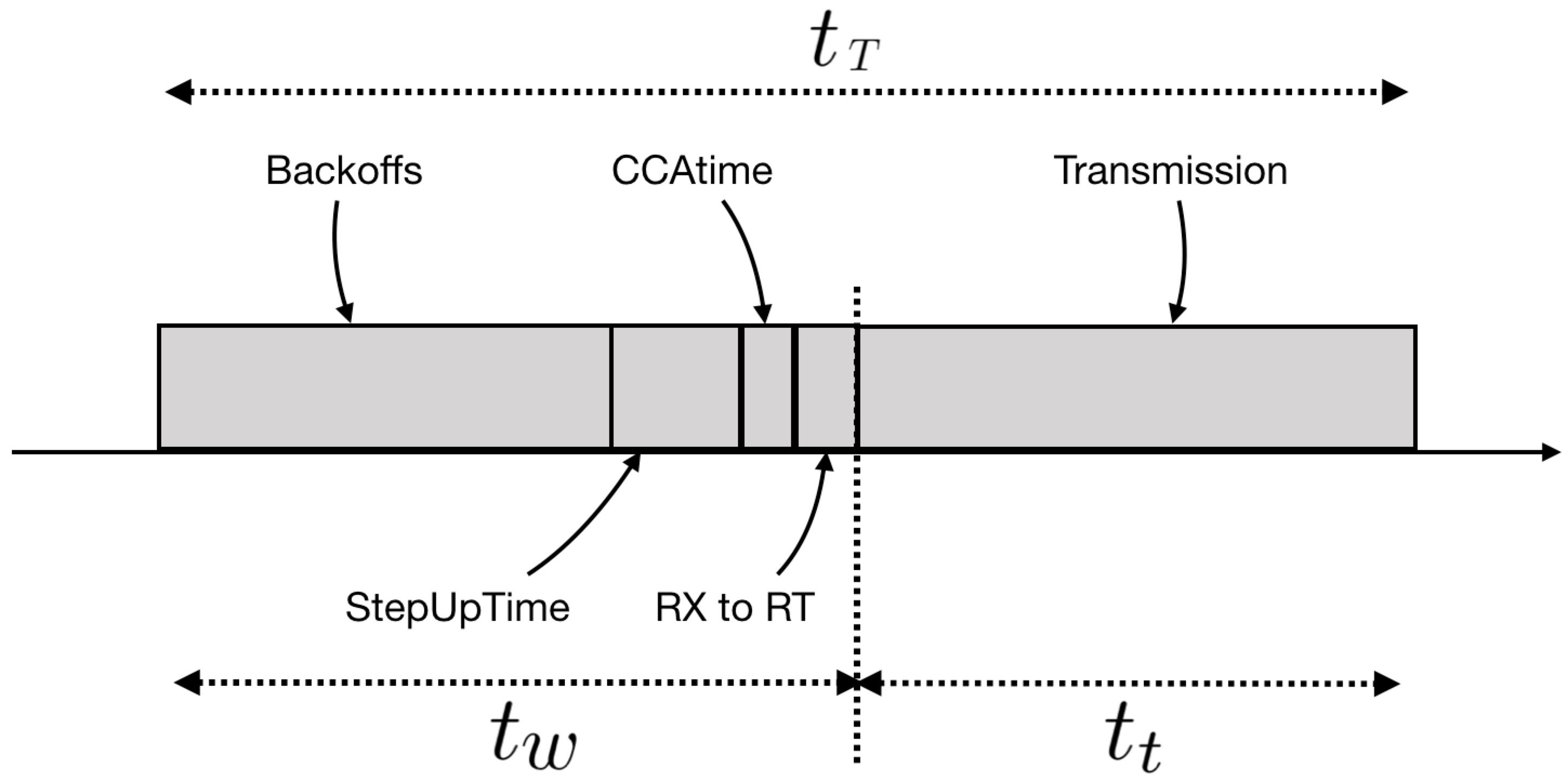}
\caption{Necessary time to complete a transmission  when CSMA/CA is used}
\label{backoff}
\end{figure}

Figure \ref{backoff} shows the time needed before finishing the transmission of a packet. When a node has a packet to send, it launches a random backoff timer according to the number of backoff units chosen randomly. After the timer expires, the node needs a SetUp time to prepare listening to the channel. A clear channel assessment time (CCAtime) is needed for a node to decide if the channel is free to transmit the packet. If the channel is free, the node then changes its state from receiver state (RX) to transmitter state (RT), and transmits the packet. If not, the node has to re-launch the backoff. All the necessary waiting time before a packet can be sent is $t_w$. The transmission time is $t_t$ and the total time needed for a packet transmission is therefore $t_T = t_w + t_t$. Note that even if CSMA/CA is used, collisions may still occur during the transmission.
%, if nodes happen to listen the channel simultaneously, then they all think the channel is idle. So they will begin to send packets simultaneously. 

\subsection{System Hypotheses}%?????????????????????????
In the sequel we assume that the system  works under the following hypothesis:
\begin{enumerate}
%\item[1] Even WBAN devices could have different appearances and sizes, we consider that all WBAN devices are represented by the same \emph{node} in the WBAN system.
%\item Nodes have enough energy to run several instances of the broadcast algorithm described in Section~\ref{System}.
\item All nodes have the same transmission power $PT$ and the same sensitivity $SN$.
%\item The transmission power can be transformed  into signal strength without loss.
\item When a node receives a packet successfully for the first time, the packet is ready to be re-transmitted with no delay.
\item Signals received with a power smaller than the sensitivity will be discarded directly by the receiver.%As some devices in the market declare their receiver sensitivity a little higher than the reality, that means even the signal strength is smaller than the declared sensitivity, packet still has chance to be detected. However, 
%\item[5] Even packets have mutual internal interference among them, the receiver will still try to receive each of them.
\item The modulation/demodulation mode in our scenario is \emph{Quadrature Phase Shift Keying}, $QPSK$.
\item We assume that the wireless channel is an \emph{Additive white Gaussian Noise} ($AWGN$) channel. The environmental noise is a Gaussian white noise, with a constant power spectral density $PN$.
\item We consider only the interferences generated by signals sent by nodes in the same WBAN that execute the broadcast protocol. That is, we assume no other protocol is executed in parallel with the broadcast protocol described above.
\end{enumerate}

\section{Markovian Model for Broadcast in WBAN}
\label{Model}

In this section we model the WBAN described in Section~\ref{System} by a continuous-time Markovian process.

\subsection{Model assumptions}

In order to derive the model we make the following assumptions:
\begin{enumerate}
\item The propagation time of the signal between two nodes within the WBAN is negligible. Considering that a WBAN is deployed on human body, this assumption is reasonable.
\item The random attenuation $A_{i,j}$ of the wireless channel between two given nodes $i$ and $j$ follows a normal distribution: $A_{i,j} \sim N(M_{i,j}, D^2_{i,j})$, where $M_{i,j}$ and $D_{i,j}$ are the mean and standard deviation of $A_{i,j}$. This assumption comes from~\cite{badreddine2015broadcast}.
\item Even though the channel attenuation fluctuates continually, we consider that during the transmission of a packet, every single bit of the considered packet experiences the same attenuation. In other words, the same value of the attenuation randomly drawn from a normal distribution will be applied to the transmission of all bits of a same packet.
\item During the reception, the Bit Error Rate ($BER$) when using QPSK as modulation in an $AWGN$ channel is calculated using the classical formula: $\frac{1}{2} \textrm{erfc} \big( \sqrt{PR/(PN+PI} \big)$, where $PR$, $PN$ and $PI$ represent the powers of the received signal, the noise spectral density and the power of the interference from other signals, respectively.
\item We assume that when several nodes simultaneously receive a packet from a given sender, the outcomes of the different receptions are independent from each other.
\item We assume that a packet has a constant size of $N_{bit}$ bits.
\end{enumerate}

\subsection{State-description}

The model focusses on the broadcast transmission of a single packet over the WBAN. As described previously, the objective of the model is to evaluate the efficiency of the multi-hop transmission. Will the considered packet be received by all nodes in the network (successful broadcast)? And if yes, after how much time?

In order to answer these questions, we consider the following state-description. At any time $t$ (not included in notations for sake of clarity), the state of the system is a $N$-tuple $(X^{0},...,X^{N-1})$, when $N$ is the number of nodes of the WBAN and $X^{i}$ is the state of node $i$ that can take three values, ``$\boldsymbol{L}$'', ``$\boldsymbol{T}$'' and ``$\boldsymbol{R}$'':
\begin{itemize}
\item[1] \textbf{Listen state $\boldsymbol{L}$}. A node in state $\boldsymbol{L}$ has not yet received the considered packet and is continuously listening waiting for its reception.
\item[2] \textbf{Transmission state $\boldsymbol{T}$}. A node in state $\boldsymbol{T}$ has correctly received the corresponding packet for the first time and is in the process to retransmit it. According to the broadcast protocol, the retransmission of the packet must be preceded by some waiting times including Backoff, listening and setup times, as described in Section 2.
\item[3] \textbf{Received state $\boldsymbol{R}$}. A node in state $\boldsymbol{R}$ has already received and retransmitted (once) the considered packet.
\end{itemize}

The normal evolution of a (non-sink) node is then to start in state $\boldsymbol{L}$ waiting for the reception of the considered packet, correctly receive it and switch to state $\boldsymbol{T}$, then run the backoff procedure and retransmit the packet, and finally switch to state $\boldsymbol{R}$. At this point, any additional reception of the same packet will not change the state of the node. If a (non-sink) node never receives correctly the considered packet, it will remain in state $\boldsymbol{L}$. Now regarding the sink node, it starts in state $\boldsymbol{T}$ and immediately after the transmission of the packet switches to state $\boldsymbol{R}$ and remains in state $\boldsymbol{R}$.

Note that when a node actually transmits a packet over the wireless channel, all other nodes are likely to listen the transmission with no delay (thanks to Assumption 1). However, this transmission can be affected by noise, as well as by attenuation between sender and receivers, and interference with other transmissions. As a result, the transmitted packet can be successfully received or unsuccessfully received by all listening nodes. According to Assumption 5, the outcomes of all possible receptions are supposed to be independent from each other.

In our scenario, $N = 7$ and sink node is node $1$. The initial state is thus $(\boldsymbol{L}^{0},\boldsymbol{T}^{1},\boldsymbol{L}^{2},\boldsymbol{L}^{3},\boldsymbol{L}^{4},\boldsymbol{L}^{5},\boldsymbol{L}^{6})$. In all remaining states of the state-description, node $1$ is in state $\boldsymbol{R}$. As a result, the maximum number of states of this description is $3^{N-1}+1$, which is equal to $730$ in our scenario.

\begin{figure}
\centering
\includegraphics[width=0.50\textwidth]{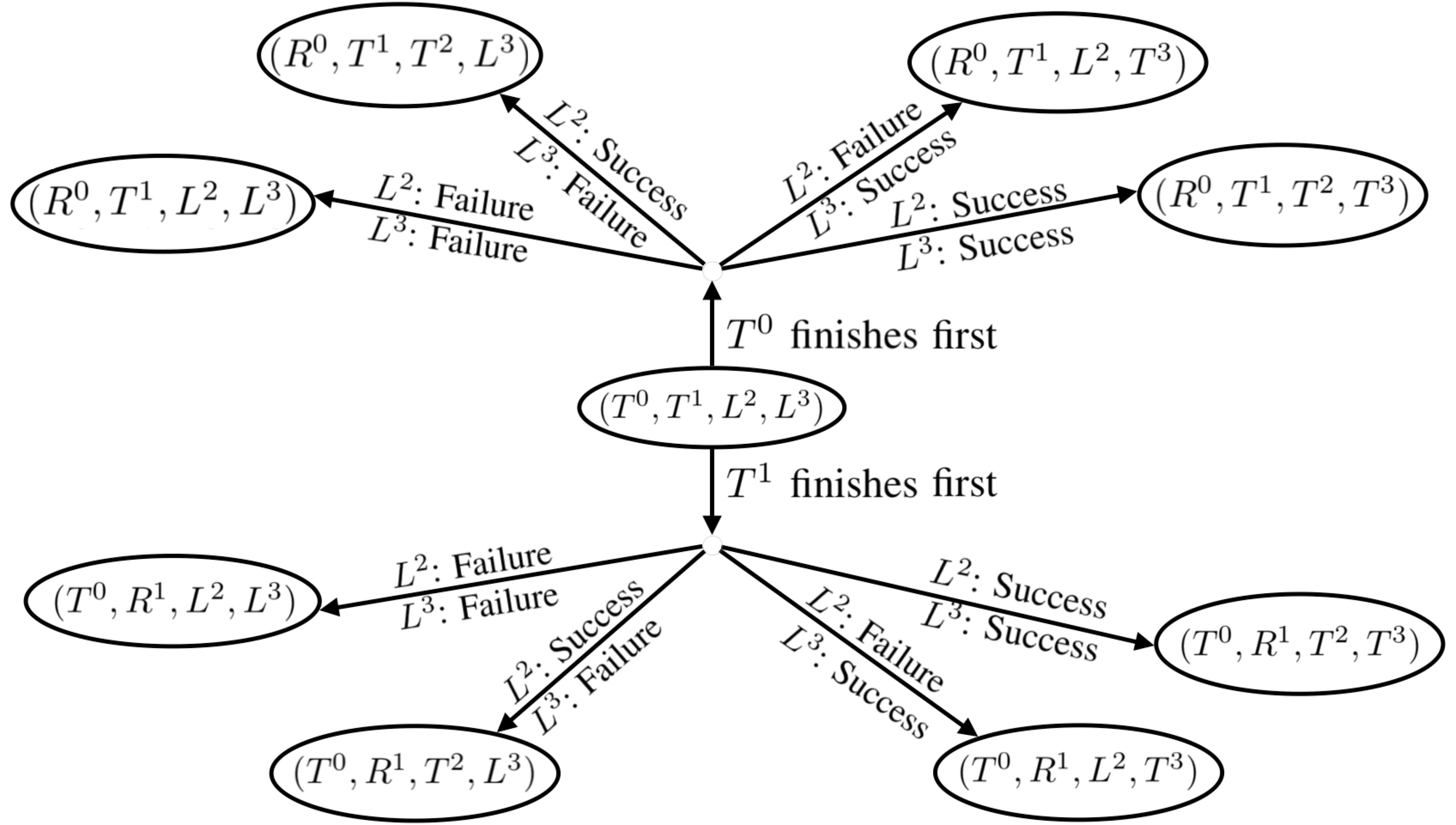}
\caption{Illustration of the state evolution in a WBAN with $N=4$ nodes.}
\label{statechange}
\end{figure}

Figure \ref{statechange} illustrates the state evolution of a WBAN made of $N=4$ nodes, starting from a given state $(\boldsymbol{T}^{0},\boldsymbol{T}^{1}, \boldsymbol{L}^{2},\boldsymbol{L}^{3})$. Because our model is continuous-time and because in this current state two nodes are in state $\boldsymbol{T}$ (nodes $0$ and $1$), only one of these two nodes will finish first its transmission of the packet. As a result, in all states that are directly accessible for the current state, one (and only one) of these two $\boldsymbol{T}$ is replaced by $\boldsymbol{R}$. Concerning the two $\boldsymbol{L}$ of the current state, each can switch to $\boldsymbol{T}$ if the corresponding node has correctly received the packet or remain in $\boldsymbol{L}$ if the reception has failed.

It is worthwhile noting that the state-diagram has no cycle. The process always starts in a predefined state (with one node in state $\boldsymbol{T}$ and all others in state $\boldsymbol{L}$) and finishes in a state where no node is anymore in state $\boldsymbol{T}$. The broadcast is a success if all nodes in the final state are in state $\boldsymbol{R}$. The broadcast is a failure if at least one node in the final state is in state $\boldsymbol{L}$.

\subsection{Markovian model}

In order to transform the evolution of this state-description into a Markovian process, we are going to assume that the time $t_T$ spent by any given node in state $\boldsymbol{T}$ is exponentially distributed with a mean $E[t_T]$, i.e., with a rate $\mu = 1/E[t_T]$. Note that, as described in Section 2, this time starts by a waiting period (backoff and setup) and ends with a transmission time. We additionally assume that this time has the same distribution for all nodes.

It thus remains to characterize the evolution of nodes in states $\boldsymbol{L}$ in a probabilistic way. To do so, if $S$ denotes the current state, e.g., $(\boldsymbol{T}^{0},\boldsymbol{T}^{1}, \boldsymbol{L}^{2},\boldsymbol{L}^{3})$ in Figure \ref{statechange}, we denote by $\mathcal{L}(S)$, $\mathcal{T}(S)$ and $\mathcal{R}(S)$, the set of nodes of $S$ that are in states $\boldsymbol{L}$, $\boldsymbol{T}$ and $\boldsymbol{R}$ respectively. For the example of Figure \ref{statechange}, $\mathcal{L}(S)=\{2,3\}$, $\mathcal{T}(S)=\{0,1\}$ and $\mathcal{R}(S)=\emptyset$. We then define $\mathcal{N}(S)$ the set of all the possible states that are directly accessible from $S$ (there are all represented in Figure \ref{statechange}). The number of states in $\mathcal{N}(S)$ is obviously:
\begin{align}
\label{Nstates}
|\mathcal{N}(S)| = |\mathcal{T}(S)| \times 2^{|\mathcal{L}(S)|} 
\end{align}

Let $U \in \mathcal{N}(S)$ be one of the possible states directly accessible from $S$. From the assumptions of independence between receptions and the supposed exponential distribution of $t_T$, we obtain the following transition rate between state $S$ and state $U$:
\begin{align}
\label{muSU}
\mu_{S, U} = |\mathcal{T}(S)| \ \mu \ \mathcal{P}_{S, U}
\end{align}
where $\mathcal{P}_{S, U}$ is given by:
\begin{align}
\label{PSU}
\mathcal{P}_{S, U} = \prod_{j\in \mathcal{L}(S) \setminus \mathcal{L}(U)}{P_{i,j}} \prod_{k\in \mathcal{L}(S) \bigcap \mathcal{L}(U)}{(1-P_{i,k})}
\end{align}
In this expression, $i$ is the (only one) node whose state changes from $\boldsymbol{T}$ to $\boldsymbol{R}$ in between $S$ and $U$, i.e., among the nodes of $S$ in state $\boldsymbol{T}$, the one that finishes its transmission first, and $P_{i,j}$ is the probability that node $j$ receives successfully the packet transmitted by $i$.

The challenge now consists in estimating probabilities $P_{i,j}$. To do so, we first develop a simple approach consisting in assuming that there is no possible interference between transmissions, leading to the so-called ``no-interference model''. Then we extend this model to take into account interferences, leading to the so-called ``general model''.

\subsection{No interference model}

Let $a_{i,j}$ be a value of the attenuation between nodes $i$ and $j$ randomly drawn from the normal distribution $A_{i,j}$ (Assumption 2), and that is applied to all bits of the considered packet under transmission by $i$ (Assumption 3). Let $PR_{i,j}$ be the power of the signal received by $j$ from $i$. Because the channel is additive, we have
\begin{align}
\label{PR}
PR_{i,j} = PT - a_{i,j}
\end{align}
where $PT$ is the transmission power (assumed to be the same for all nodes). This signal will be strong enough to be heard by node $j$ if the power of the received signal is above a given threshold $SN$. This implies that the first condition for $j$ to be able to receive the signal sent by $i$ is that $a_{i,j}$ must be less than $a_{max} = PT-SN$.

Now, according to Assumption 4, the BER experienced by all bits of the considered packet is:
\begin{align}
\label{BER}
BER = \frac{1}{2} \textrm{erfc} \Big( \sqrt{PR_{i,j}/(PN+PI_{i,j}}) \Big)
\end{align}
where  $PI_{i,j}$ corresponds to the power of all interfering signals, i.e., signals issued from all nodes that are transmitting simultaneously with node $i$.

In this first ``no-interference model'', we neglect the influence of interference between simultaneous transmissions and thus assume that $PI_{i,j} = 0$. The $BER$ thus simplifies as:
\begin{align}
\label{BERno}
BER_{no} = \frac{1}{2} \textrm{erfc} \Big( \sqrt{PR_{i,j}/PN} \Big)
\end{align}

Conditioned by the fact that the the signal is strong enough to be heard by the receiver, the probability that it can be decoded successfully is therefore:
\begin{align}
\label{Pdno}
Pd_{no} = (1-BER_{no})^{N_{bit}}
\end{align}

We can then calculate the probability $P_{i,j}(a_{i,j})$ that node $j$ receives successfully the packet transmitted by $i$ conditioned by the fact that the attenuation of the signal is $a_{i,j}$:
\begin{align}
\label{Pija}
P_{i,j}(a_{i,j}) =
\begin{cases}
	Pd_{no}, &\text{if} \quad a_{i,j} < a_{max}\\
	0, &\text{otherwise}
\end{cases}
\end{align}

In order to finally derive the probabilities $P_{i,j}$ that appear in the expression of the transitions rates $\mu_{S, U}$ (relations \ref{muSU} and \ref{PSU}), we just have to decondition $P_{i,j}(a_{i,j})$ with regard to the distribution of $A_{i,j}$:
\begin{align}
\label{Pij}
\mathcal{P}_{i,j} = \int_{0}^{\infty} P_{i,j}(x) f_{A_{i,j}}(x) dx
\end{align}
where $f_{A_{i,j}}$ is the PDF of the normal distribution $A_{i,j}$.

\subsection{General model}

We now introduce interferences in the model. An interference occurs when two nodes (or more) in the WBAN simultaneously transmit a packet, and another node hears both transmissions. In such a case, we will assume that the transmission that has begun first is the one that the receiving node will try to decode, and will be referred to as the ``target signal''. The other transmission (or the others in case of multiple interferences) will be considered as ``interfering signal''. The target signal can still be successfully decoded by the receiving node if the interfering signal is not too strong with regards to the target signal and if the overlap between the target signal and the interfering signal is not too long.

Let $S$ be the current state of the Markovian process, and let $U \in \mathcal{N}(S)$ be a possible ``next state'', i.e., a state directly accessible from $S$. As in the previous subsection, we define $i$ as the (only one) node whose state changes from $\boldsymbol{T}$ to $\boldsymbol{R}$ in between $S$ and $U$, i.e., the node in $S$ that finishes its transmission first. The signal sent by $i$ will thus be the target signal for all listening nodes in $S$ that can hear it. Indeed, because a packet has a fixed size, its transmission lasts for a constant time, and as a result, the first transmission that starts is the first that ends. We then define $\mathcal{T}_i(S) = \mathcal{T}(S) \setminus \{i\}$ the set of nodes in $S$ that may interfere with node $i$.

\subsubsection{Single interfering node}

\begin{figure}
\centering
\includegraphics[width=0.48\textwidth]{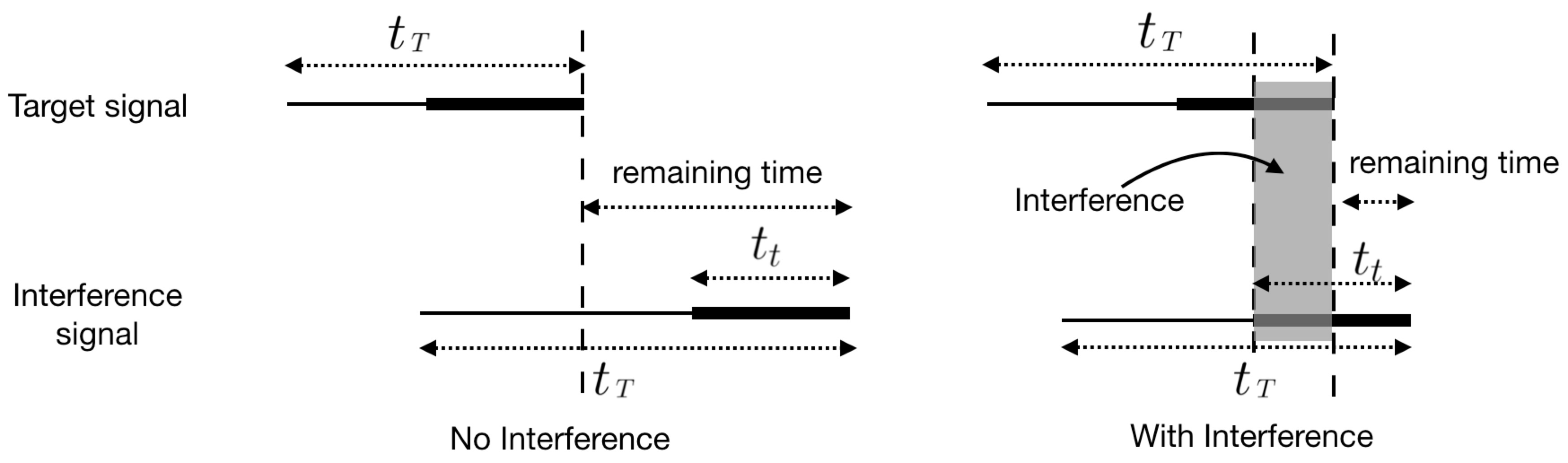}
\caption{No-Interference case vs With-Interference case}
\label{inE}
\end{figure}

Let us first assume that $\mathcal{T}_i(S)$ is made of a single node labeled as $k$, i.e., $\mathcal{T}_i(S) = \{k\}$. The target signal is issued from node $i$ and the interfering signal is issued from node $k$.  Knowing that $i$ is the first to finish its transmission, the transmission of $k$ will interfere with the transmission of $i$ only if the remaining time node $k$ stays in state $\boldsymbol{T}$ is greater than $\bar{t}_t$ (see Figure~\ref{inE} for illustration). Remember that we assume that the time $t_T$ spent by any node in state $\boldsymbol{T}$ is exponentially distributed. Because of the memoryless property of the exponential distribution, the probability that node $k$ interfere with node $i$ is given by:
\begin{align}
\label{pI}
p_I = P\{t_T < \bar{t}_t\} = F_{t_T}(\bar{t}_t)
\end{align}
where $F_{t_T}$ is the cumulative distribution of $t_T$:
\begin{align}
\label{FtT}
F_{t_T}(t) = 1-e^{-\mu t} = 1-e^{-\frac{t}{E[t_T]}}
\end{align}

Conditioned by the fact that node $k$ does not interfere with node $i$ (i.e., with a probability $1-p_I$), the BER experienced by all bits of the transmission of node $i$ is given by relation~\ref{BERno}. And if the signal is strong enough to be heard by the receiver, the probability $Pd_{no}$ that it can be decoded successfully is given by relation~\ref{Pdno}. 

Now, conditioned by the fact that node $k$ interfere with node $i$ (i.e., with a probability $p_I$), in average half of the bits of the transmission experience a BER without interference (given by relation~\ref{BERno}) and half of the bits experience a BER with interference given by:
\begin{align}
\label{BERint}
BER_{int} = \frac{1}{2} \textrm{erfc} \Big( \sqrt{PR_{i,j}/(PN+PI_{i,j}}) \Big)
\end{align}
where the interfering power $PI_{i,j}$ here corresponds to the signal issued from node $k$: 
\begin{align}
\label{PI}
PI_{i,j} = PT - E[A_{k,j}]
\end{align}
Note that we approximately use the mean of the (normal) random variable $A_{k,j}$ in the expression of the interfering power. Now, if the signal is strong enough to be heard by the receiver, the probability $Pd_{int}$ that it can be decoded successfully is:
\begin{align}
\label{pdint}
Pd_{int} = (1-BER_{int})^{\frac{N_{bit}}{2}} \times (1-BER_{no})^{\frac{N_{bit}}{2}}
\end{align}

Finally the probability $Pd$ that the packet can be successfully decoded can be expressed as:
\begin{align}
\label{pd}
Pd = (1-p_I) Pd_{no} + p_I Pd_{int}
\end{align}

We can finally calculate the probability $P_{i,j}(a_{i,j})$ that node $j$ receives successfully the packet transmitted by $i$ conditioned by the fact that the attenuation of the signal is $a_{i,j}$, in the same way as in relation~\ref{Pija} (by replacing $Pd_{no}$ by $Pd$), and decondition this expression by using relation~\ref{Pij}.

\subsubsection{Multiple interfering nodes}

We now consider the general case where $\mathcal{T}_i(S)$ is made of a several nodes. Every node in $\mathcal{T}_i(S)$ is likely to interfere with the transmission of node $i$ (which is assumed to be the first one to finish). Let $P(\mathcal{T}_i(S))$ be the powerset of $\mathcal{T}_i(S)$, i.e., the set of all possible subsets of $\mathcal{T}_i(S)$. Any subset $\mathcal{X} \in P(\mathcal{T}_i(S))$ will correspond to a set of nodes that actually interfere with node $i$, i.e., nodes which transmission overlaps with the transmission of node $i$. This situation occurs with a probability given by:
\begin{align}
\label{gongshi13}
p(\mathcal{X}) = p^{|\mathcal{X}|}_I \times (1-p_I)^{|\mathcal{T}_i(S)| - |\mathcal{X}|}
\end{align}
where $p_I$ is given by relation~\ref{pI}. Note that $p(\emptyset) = (1-p_I)^{|\mathcal{T}_i(S)|}$ corresponds to the probability that there is no interfering node.

We then propose the following approximation for the probability $Pd$ that the packet can be successfully decoded:
\begin{align}
\label{pd2}
Pd = p(\emptyset) Pd_{no} + \sum_{\mathcal{X} \in P(\mathcal{T}_i(S)) \setminus \emptyset}{p(\mathcal{X}) Pd_{int}(\mathcal{X})}
\end{align}
where $Pd_{int}(\mathcal{X})$ has exactly the same expression as $Pd_{int}$ in relation~\ref{pdint}, but with a BER that is obtained from relation~\ref{BERint} with an interfering power $PI_{i,j}$ estimated as:
\begin{align}
PI_{i,j} = \sum_{k \in \mathcal{X}}{(PT - E[A_{k,j}])}
\end{align}

\section{Numerical results}
\label{NumRes}
In this section, we first validate our model by comparison with simulation. For that purpose, we used the open source networking simulator $OMNeT++$ (version 4.6) combined with the framework $MIXIM$ that provides the CSMA/CA protocol utilized in WBANs. We borrowed the wireless channel implementation from \cite{badreddine2015broadcast} (additive channel and random attenuation) based on data sets from \cite{naganawa2015simulation}. The means and standard deviations of the random attenuation of all channels between each pair of nodes in posture $2$ are shown in Table \ref{dataSet2}. Values above the diagonal are the means values and values above the diagonal are the standard deviation.

\begin{table}[!hbp]
\centering
\caption{Means and Standards Deviations of Path Loss for all the links in Posture 2) Running \cite{naganawa2015simulation}}
\label{dataSet2}
\resizebox{.5\textwidth}{!}{
\begin{tabular}{c|ccccccc|c}
\hline
 \huge{$T_{X}$ or $R_{X}$} & \huge{navel} & \huge{chest} & \huge{head} & \huge{upper arm} & \huge{ankle} & \huge{thigh} & \huge{wrist} & \\ \hline
\huge{navel} & \diagbox[width=6em,trim=l] & \huge{31.4} & \huge{47.4} & \huge{54.5} & \huge{57.9} & \huge{44.8} & \huge{45.9} &  \\
\huge{chest} & \huge{1.4} & \diagbox[width=6em,trim=l] & \huge{41.0} & \huge{39.2} & \huge{61.0} & \huge{49.9} & \huge{41.2} &  \\
\huge{head} & \huge{3.5} & \huge{2.9} & \diagbox[width=6em,trim=l] & \huge{41.3} & \huge{65.6} & \huge{59.3} & \huge{45.5} &  \\
\huge{upper arm} & \huge{9.9} & \huge{8.4} & \huge{8.4} & \diagbox[width=6em,trim=l] & \huge{58.0} & \huge{52.4} & \huge{33.8} & \huge{Mean[dB]} \\
\huge{ankle} & \huge{6.9} & \huge{6.9} & \huge{5.7} & \huge{8.2} & \diagbox[width=6em,trim=l] & \huge{39.0} & \huge{56.9} &  \\
\huge{thigh} & \huge{2.2} & \huge{4.8} & \huge{7.3} & \huge{7.8} & \huge{1.8} & \diagbox[width=6em,trim=l] & \huge{49.6} &  \\
\huge{wrist} & \huge{6.1} & \huge{8.2} & \huge{3.5} & \huge{4.6} & \huge{7.5} & \huge{11.6} & \diagbox[width=6em,trim=l] &  \\ \hline
 &  &  &  \multicolumn{3}{c}{\huge{Standard deviation [dB]}}  &  &  & 
\end{tabular}
}
\end{table}

For the network layer we implemented the broadcast protocol described in Section \ref{System}.
All the performance parameters presented in this section are plotted as a function of the transmission power ranging from $-60$ dBm to $-50$ dBm. They all correspond to posture $2)$ Running, which is a high mobility posture.
All curves present the comparison between the ``no-interference model'', the ``general model'' (described in Section~\ref{Model}) and simulation. Each simulation point is the average of 1000 executions. 

In a second part, we propose a ``Multi-Broadcast'' protocol that aims at improving the broadcast efficiency while using a low transmission power. We use our validated model to show the trade-off between the number of runs of the multi-broadcast procedure and the transmission power.

\subsection{Validation}

\subsubsection{Cover Probability}
The cover probability is the probability that the outcome of the broadcast is a success, i.e., the probability that all (non-sink) nodes in the WBAN eventually receive the considered packet successfully. In the model, this performance index is evaluated by the probability, starting from the initial state $I = (\boldsymbol{L}^{0}$, $\boldsymbol{T}^{1}$, $\boldsymbol{L}^{2}$, $\boldsymbol{L}^{3}$, $\boldsymbol{L}^{4}$, $\boldsymbol{L}^{5}$, $\boldsymbol{L}^{6})$ to reach the final state $F = (\boldsymbol{R}^{0}$, $\boldsymbol{R}^{1}$, $\boldsymbol{R}^{2}$, $\boldsymbol{R}^{3}$, $\boldsymbol{R}^{4}$, $\boldsymbol{R}^{5}$, $\boldsymbol{R}^{6})$.

\begin{figure}
\centering
\includegraphics[width=0.45\textwidth]{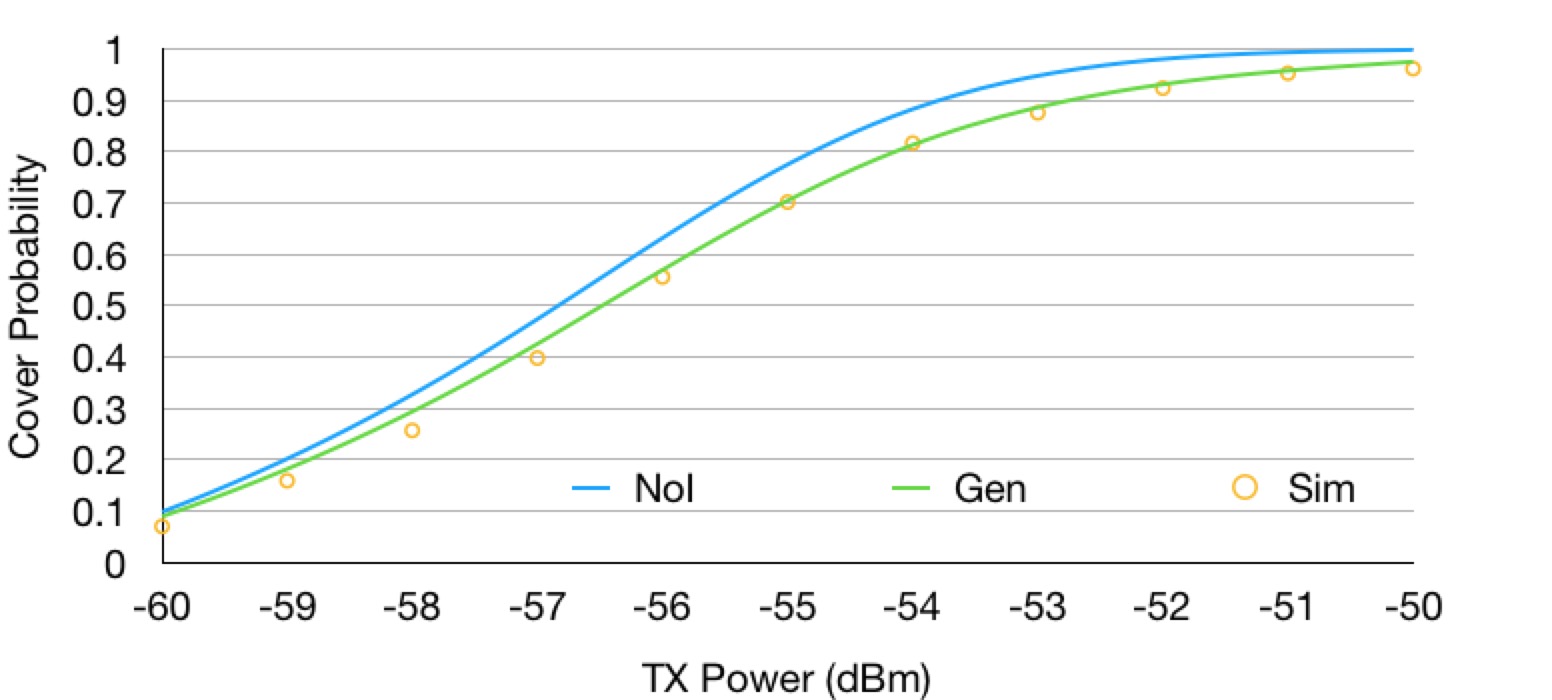}
\caption{Cover Probability}
\label{coverprobability}
\end{figure}

Figure \ref{coverprobability} shows the cover probability evaluated by both models (no-interference and general) and by simulation. Obviously the cover probability is an increasing function of the transmission power. As shown by the figure, the general model gives better results than the simpler no-interference model, when we compare them to simulation. Indeed, the average relative error (between model and simulation) is around $14\%$ for the no-interference model and is less than $6\%$ for the general model.

\subsubsection{Average Cover Number}
The average cover number represents the average number of (non-sink) nodes in the WBAN that receive successfully the considered packet. This performance parameter is complementary to the cover probability. When the cover probability tends to $1$, the average cover number obviously tends to $6$ (not including the sink node). However, for lower values of the cover probability, the average cover number gives more information about the number of nodes that are responsible of the failure of the broadcast.
In the model, this performance parameter is evaluated as the average number of nodes in states $\boldsymbol{R}$ in any possible final state (reachable from initial state $I$) of the Marvov Chain evolution.

\begin{figure}
\centering
\includegraphics[width=0.45\textwidth]{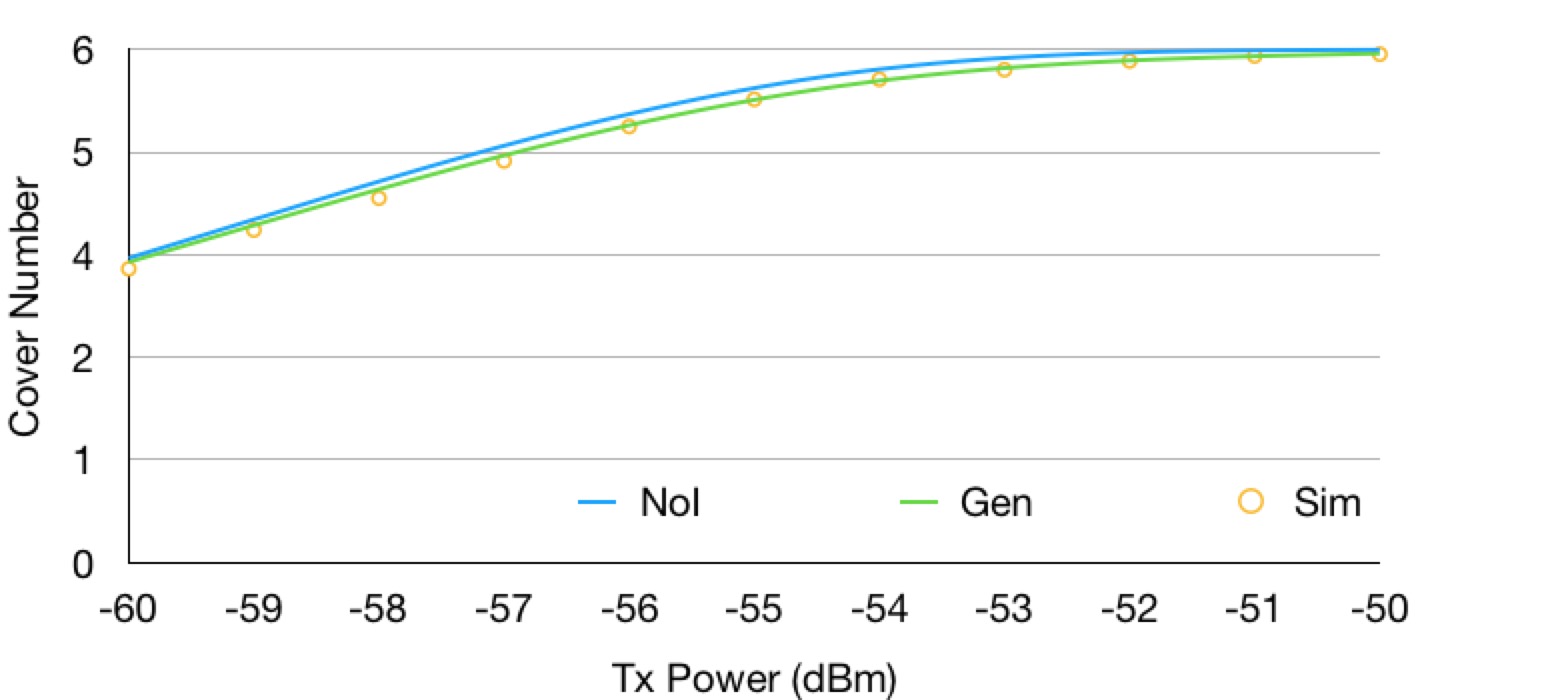}
\caption{Average Cover Number}
\label{covernumber}
\end{figure}

As shown in Figure~\ref{covernumber}, both models give results that are very close to simulation, but still the general model remains better than the no-interference model.

\subsubsection{Hitting Probability}
The hitting probability is the probability that a particular node (chosen among all non-sink nodes) successfully receive the considered packet. As a result, we are not evaluating the outcome of the complete broadcast anymore, but we concentrate on the successful reception of the packet by a particular node, but still when using the broadcast protocol (and not a unicast transmission).
In the model, this performance parameter corresponds to the probability of reaching any final state in which the particularized node is in state $\boldsymbol{R}$.

\begin{figure}
\centering
\includegraphics[width=0.45\textwidth]{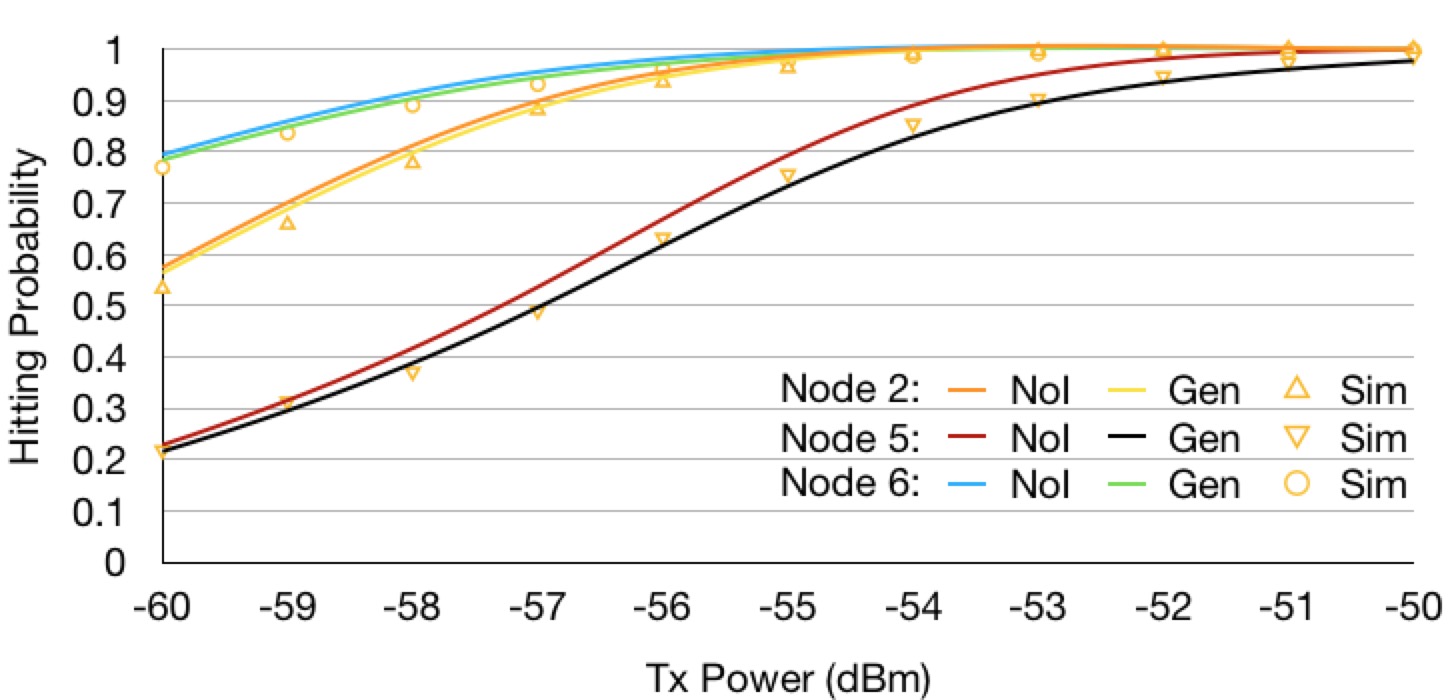}
\caption{Hitting Probability for nodes 2, 5 and 6}
\label{hittingprobability}
\end{figure}

Figure \ref{hittingprobability} gives three sets of curves, each corresponding to a given particularized node: node 2) head, node 5) thigh and node 6) wrist. Once again, we see that both models give accurate results and that the general model produces better estimations.

\subsubsection{Average Cover Time}
The average cover time corresponds to the mean duration of the overall broadcast procedure, i.e., the average time that is necessary for all (non-sink) nodes to receive successfully the considered packet, conditioned by the fact that the broadcast is a success. Indeed, if the broadcast fails, this parameter has no interest.
We take advantage of the fact that our Markovian model is continuous-time to estimate this performance parameter as the average time spent in all the paths of the Markov chain that start in the initial state $I$ and ends in the final state $F$.

\begin{figure}
\centering
\includegraphics[width=0.45\textwidth]{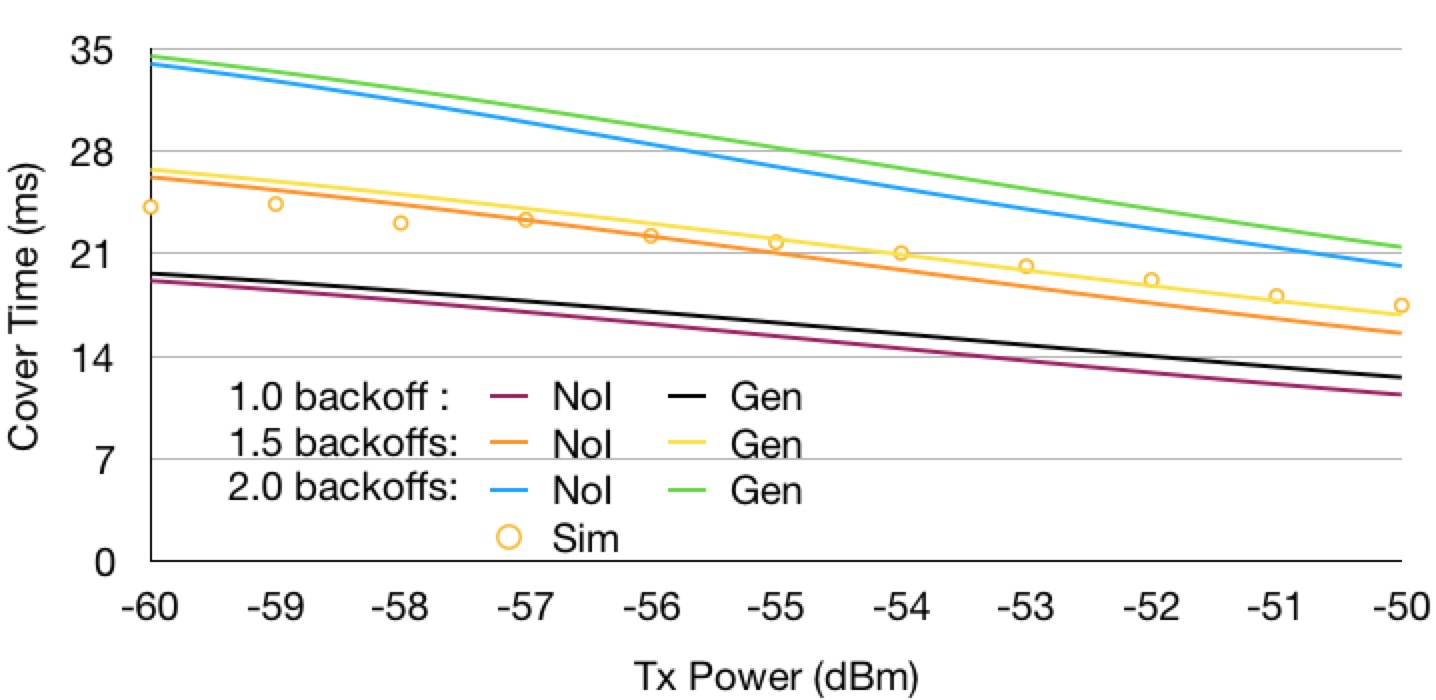}
\caption{Average Cover Time}
\label{covertime}
\end{figure}

However, in order to be able to calculate the time spent in a given path of the chain, we need to know the average time spent in each state of the path. From relation~\ref{muSU}, the average time spent in a given state $S$ of the Markov chain is $\frac{1}{|\mathcal{T}(S)| \ \mu} = \frac{E[t_T]}{|\mathcal{T}(S)|}$. We thus need to have an estimation of $E[t_T]$. Remember that $t_T$ is the time spent by any node in state $\boldsymbol{T}$ (supposed to be exponentially distributed), this time depending mainly on the duration of the backoff preceding the transmission. This duration is related to the average number of backoff periods that precede the transmission. In Figure \ref{covertime} we give the average cover time corresponding to three different values for the average number of backoff periods: $1$, $1.5$ and $2$. As can be seen on the corresponding curves, $1$ leads to an under-estimation of the average cover time, $2$ leads to an over-estimation, and $1.5$ enables to obtain very accurate results.

As a result, the average cover time can be accurately estimated provided that we can estimate the average number of backoff periods preceding the transmission of a node. This parameter depends on the actual configuration (posture, attenuation, transmission power, etc.) as well as on the load of the system. The derivation of this parameter would need to develop a fixed-point iteration method (as it depends on the performance metrics that in turn depend on its estimated value) and is left for future extensions.

\subsection{Multi-broadcast protocol}

As shown by figures~\ref{coverprobability} and \ref{covernumber}, the efficiency of the broadcast procedure can be improved by increasing the transmission power of all nodes. However, because a WBAN is deployed on human body, the transmission power may be limited for health reasons. We then propose a ``Multi-Broadcast'' procedure that aims at increasing the efficiency of the broadcast without increasing the transmission power, by simply broadcasting several times the same packet.

We associate with the multi-broadcast a parameter $K$ that corresponds to the number of times a same packet has to be successively broadcasted. Each of the $K$ attempts for broadcasting a given packet will increase the probability that the packet will be received at least once by all (non-sink) nodes. In order to evaluate the performance of the multi-broadcast protocol, we can use either the simple no-interference model or the more precise general model. Note that contrarily to simulation that needs to run the $K$ successive broadcasts, the model has to be run only once. Indeed, the performance parameters of the $K$-multi-broadcast can be easily derived from the performance parameter of the single-broadcast model. As an example, the cover probability for the $K$-multi-broadcast will be derived from the Markov chain of the single-broadcast, by calculating the probability of reaching $K$ final states (one for each broadcast), that are such that each (non-sink) node is in state $\boldsymbol{R}$ in at least one of the $K$ final states.

\begin{figure}
\centering
\includegraphics[width=0.45\textwidth]{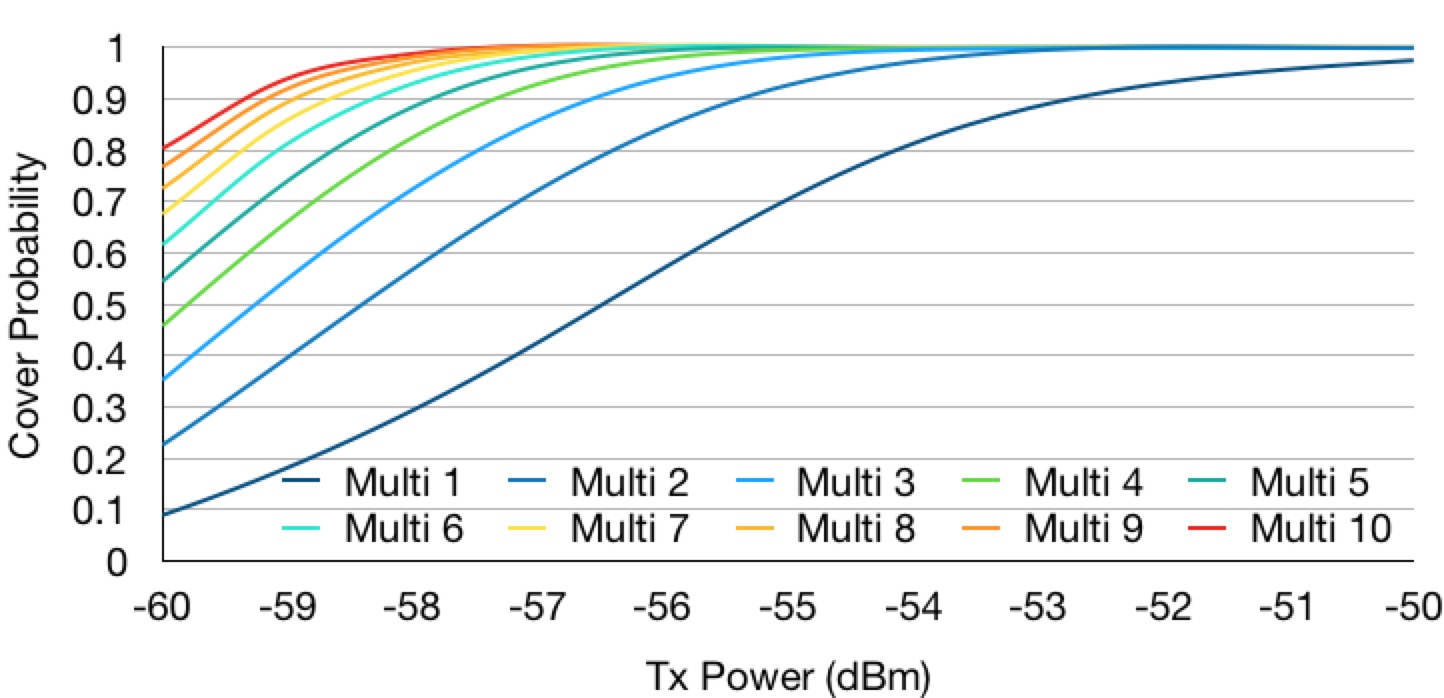}
\caption{Cover probability for the Multi-Broadcast protocol}
\label{MultiB}
\end{figure}

Figure \ref{MultiB} presents the cover probability for the $K$-multi-broadcast, for $K$ ranging between $1$ (equivalent to the single-broadcast case) and $10$, when using the general model. The trade-off between using greater transmission powers or greater values of $K$ clearly appears. For example, to keep a cover probability above $90\%$, we can either run a single-broadcast with a transmission power of $-52.5$ dBm or run a $4$-multi-broadcast with a transmission power of $-57.5$ dBm. The model thus enables to derive Erlang-like abaques in order to dimension the parameter $K$ of the proposed multi-broadcast protocol.

\section{Conclusion}
%In this paper, we modelling a medical application dedicated network, WBAN, working on a broadcast scenario. We propose both a general model where take the intern interference into account and a no-interference model, an approximation of the general model. We then validated our model by comparing with realistic simulation. A further application, multi-broadcast, is also evaluated by our model.
%In this paper, we  presented an analytical model to evaluate multi-hop broadcast protocols in Wireless Body Area Networks.  We validate our model by evaluating the  efficiency of a case study protocol (a diffusion-based broadcast), e.g., its cover probability, and to help in dimensioning its parameters, e.g., the transmission power of sensors. We have compared the performance derived from our model to OMNeT++ simulations and shown its accuracy. Lastly, in order to provide a highly reliable broadcast while keeping low transmission power, we proposed a multi-broadcast protocol that consists in broadcasting several times the same packet. The simplicity of our model  allowed us to evaluate the approach and to show the trade-off between transmission power and redundancy. As future work we aim at  adapting our model to other communication protocols in WBANs, such as convergecast.
%
In this paper, we have presented an analytical model to evaluate broadcast protocol in Wireless Body Area Networks. In order to reduce the transmission power of body sensors, a multi-hop approach has been developed, where sensors act as relays for broadcasting some information. The proposed model enables to evaluate the efficiency of the protocol, e.g., the cover probability, and to help in dimensioning its parameters, e.g., the transmission power of sensors. We have compared the performance derived from our model to OMNeT++ simulations and shown its accuracy. Lastly, in order to provide a highly reliable broadcast while keeping low transmission power, we have proposed a multi-broadcast protocol that consists in broadcasting several times the same packet. The simplicity of our model has allowed us to evaluate the approach and to show the trade-off between transmission power and redundancy. In future work, we aim at showing that, even though our model is dedicated to broadcast, it can be easily adapted to other communication protocols in WBANs, such as the convergecast of data collection.  

\bibliographystyle{plain}
\bibliography{sample}
\end{document}